\documentclass[12pt]{article}
\def\AP#1#2#3{{Ann.\ Phys.} {#1}, #2 (#3)}

\def\PRD#1#2#3{{Phys.\ Rev.} {D#1}, #2 (#3)}

\def\NPB#1#2#3{{Nucl.\ Phys.} {B#1}, #2 (#3)}

\def\PLB#1#2#3{{Phys.\ Lett.} {B#1}, #2 (#3)}

\def\IJMPA#1#2#3{{Int.\ J.\ Mod.\ Phys.} {A#1}, #2 (#3)}

\def\MPLA#1#2#3{{Mod.\ Phys.\ Lett.} {A#1}, #2 (#3)}
\def\CQG#1#2#3{{Class.\ Quantum Grav.} {#1}, #2 (#3)}

\def\ANP#1#2#3{{Annalen Phys.} {#1}, #2 (#3)}

\def\GC#1#2#3{{Grav.\ \& Cosm.} {#1}, #2 (#3)}
\newcommand{\be}{\begin{equation}}
\newcommand{\ee}{\end{equation}}
\newcommand{\ba}{\begin{eqnarray}}
\newcommand{\ea}{\end{eqnarray}}
\begin{document}
\title{A note on Weyl transformations in two-dimensional dilaton
gravity\thanks{This work is supported in part by funds provided by the U.S.\
Department of Energy (D.O.E.) under cooperative research agreement
DE-FC02-94ER40818, and by a FCT grant, contract number BPD/20166/99.}}
\author{Marco Cavagli\`a
\thanks{Post office address:
Massachusetts Institute of Technology,
Marco Cavagli\`a 6-408,
77\hfill\break Massachusetts Avenue,
Cambridge MA 02139-4307, USA.}
\thanks{E-mail: cavaglia@mitlns.mit.edu}\\
\small Center for Theoretical Physics\\
\small Laboratory for Nuclear Science and Department of Physics\\
\small Massachusetts Institute of Technology\\
\small 77 Massachusetts Avenue, Cambridge Massachusetts 02139, USA\\
\small and\\
\small Istituto Nazionale di Fisica Nucleare, Sede di Presidenza, Roma, Italia.}
\date{(MIT-CTP-3047, hep-th/0011136. \today)}
\maketitle

\begin{abstract}
We discuss Weyl (conformal) transformations in two-dimensional matterless
dilaton gravity. We argue that both classical and quantum dilaton gravity
theories are invariant under Weyl transformations.
\end{abstract}

\leftline{PACS number(s): 04.60.Kz, 04.20.Cv}

\section*{Introduction}
This letter deals with Weyl transformations in two-dimensional dilaton gravity.
Motivated by some recent papers about the role of Weyl transformations in
two-dimensional dilaton gravity (see e.g.\ \cite{GHK,KV}), we want to answer
the following question: Is two-dimensional matterless dilaton gravity invariant
under Weyl conformal transformations? A short answer to this question seems to
be negative \cite{GHK,KV}. However, a more careful analysis shows that the {\it
physical} properties of classical and quantum dilaton gravity theories are
actually invariant under Weyl rescalings of the metric. This paper provides
evidence in support of this claim. 
\section*{The curse of Weyl transformations}
Our starting point is the action
\be
S_{DG}=\int_\Sigma d^2x\sqrt{-\gamma}\,\left[\phi
R^{(2)}(\gamma)+{\cal V}(\phi)-{d~\over d\phi}[\ln|{\cal W}(\phi)|]
(\nabla\phi)^2\right]\,,\label{l1}
\ee
where ${\cal V}(\phi)$ and ${\cal W}(\phi)$ are functions of the dilaton $\phi$
and $\gamma_{\mu\nu}$ is a two-dimensional metric with hyperbolic signature.
The interest in the two-dimensional dilaton gravity theories (\ref{l1}) is
motivated by their relation to a number of (physical) $N$-dimensional
spacetimes, such as black holes and p-branes, via compactification of $N-2$
dimensions. The most remarkable example is the $N$-dimensional spherically
symmetric black hole which is described by an effective two-dimensional theory
(\ref{l1}) upon dimensional reduction by integration on the spherical
coordinates \cite{Cavaglia-PRD,CD-GC}.

As is well-known \cite{BL,LGK}, a classical Weyl transformation 
\be
\gamma_{\mu\nu}(x)=\tilde\gamma_{\mu\nu}(x)\,{1\over\Omega(\phi)}\,,\label{l2}
\ee
where $\Omega(\phi)$ is a generic function of the dilaton, is often used to
simplify Eq.\ (\ref{l1}). Choosing $\Omega(\phi)=1/{\cal W}(\phi)$ the  kinetic
term of the dilaton in the action (\ref{l1}) can be set to zero. The action
becomes
\be
\tilde S_{DG}=\int_\Sigma d^2x\sqrt{-\tilde\gamma}\,\left[\phi
\tilde R^{(2)}(\tilde\gamma)+\tilde{\cal V}(\phi)\right]\,,\label{l3}
\ee
where
\be
\tilde{\cal V}(\phi)={{\cal V}(\phi)\over\Omega(\phi)}=
{\cal V}(\phi){\cal W}(\phi)\,.
\label{l4}
\ee
Many authors consider Weyl transformations to be legitimate in classical
dilaton gravity theories \cite{BL,LGK}. However, some concerns have been raised
about their use in quantum theory \cite{GHK,KV}. The main objection to using
Weyl transformations (\ref{l2}) is that Weyl-related theories may describe
locally inequivalent theories and change the global structure of the theory. An
example which is frequently found in the literature is dilaton gravity with
constant dilatonic potential in the twiddle frame (\ref{l3}), also known as the
Callan-Giddings-Harvey-Strominger (CGHS) matterless model \cite{CGHS}
\be
\tilde S_{CGHS}=\int_\Sigma d^2x\sqrt{-\tilde\gamma}\,[\phi
\tilde R^{(2)}(\tilde\gamma)+4\lambda^2]\,.\label{l5}
\ee
In the untwiddle frame the CGHS model is described by Eq.\ (\ref{l1}) with
${\cal V}(\phi)=4\lambda^2\phi$ and ${\cal W}(\phi)=\phi^{-1}$. The two frames
are related by the  Weyl transformation (\ref{l2}) with $\Omega(\phi)=\phi$.
Varying Eq.\ (\ref{l5}) with respect to $\phi$ it is straightforward to prove
that the CGHS model in the twiddle frame describes a flat Minkowski spacetime.
This is not true in the untwiddle frame. So both local properties and global
structure of the spacetime in the two frames are different. This fact has often
been used to support the claim that models that are related by Weyl
transformations describe, generally, dynamically inequivalent theories.
\section*{The rescue of Weyl transformations}
The conclusion that we have reached in the previous section is inaccurate, both
from classical and quantum points of view.

Let us consider first the classical theory. While is correct to say that the
twiddle and untwiddle theories are dynamically nonequivalent when considered
separately, the two theories have the same classical, i.e., {\it physical},
content {\it if} the Weyl transformation (\ref{l2}) which relates the two
frames is properly taken into account. We have the following

\medskip\noindent
{\bf Theorem.} Classical two-dimensional pure dilaton gravity theory is
invariant under the (Weyl) transformation
\be
{\cal V}(\phi)=\tilde{\cal V}(\phi)\Omega(\phi)\,,\quad
{\cal W}(\phi)=\tilde{\cal W}(\phi)/\Omega(\phi)\,,\quad
\gamma_{\mu\nu}=\tilde\gamma_{\mu\nu}/\Omega(\phi)\,.
\label{l6}
\ee
where $\Omega(\phi)$ is an arbitrary function of the dilaton.

\medskip\noindent {\bf Proof.} Consider the action (\ref{l1}). Under the transformation
(\ref{l6}) the Lagrangian transforms as 
\be
{\cal L}=\tilde{\cal L}+\sqrt{-\tilde\gamma}\tilde\nabla\left[\phi
\tilde\nabla(\ln|\Omega|)\right]\,.
\label{l7}
\ee
Therefore, the transformation (\ref{l6}) does not affect the equations of
motion and is a symmetry of the model. This property can also be directly
checked by implementing the transformation (\ref{l6}) into the equations of
motion
\ba
\nabla_{(\mu}\nabla_{\nu)}\phi&-&g_{\mu\nu}\nabla^2\phi
+{1\over 2}g_{\mu\nu}V(\phi)~+\nonumber\\
&+&\displaystyle {d~\over d\phi}\ln|{\cal W}(\phi)|
[\nabla_{(\mu}\phi\nabla_{\nu)}\phi-{1\over 2}g_{\mu\nu}(\nabla\phi)^2]=0\,,
\label{l8}\\
R^{(2)}(g)&+&2\nabla^2 \ln|{\cal W}(\phi)|~-~
(\nabla\phi)^2{d^2~\over d\phi^2}\ln|{\cal W}(\phi)|
~+~{d~\over d\phi}{\cal V}(\phi)=0\,.
\label{l9}
\ea
It is straightforward to check that the transformation (\ref{l6}) leaves
invariant Eq.\ (\ref{l8}) and Eq.\ (\ref{l9}). [Hint: to prove the invariance
of Eq.\ (\ref{l9}) use the trace of Eq.\ (\ref{l8}).] 

So the two theories are physically equivalent at classical level. Using the
invariance (\ref{l6}) one can always set the dilatonic potential ${\cal
V}(\phi)$ or the dilatonic kinetic coupling ${\cal W}(\phi)$ to a given
function. The choice of $\Omega(\phi)$ coincides with a ``gauge fixing'' for
the symmetry, Eq.\ (\ref{l6}). Clearly, the local properties of the gauge-fixed
metric (e.g.\ the curvature) generally depend on the particular gauge that has
been chosen.  However, only quantities that are invariant under the symmetry
(\ref{l6}) should be considered. It is particularly instructive to discuss this
point in the context of dimensionally reduced models. Let us consider the
$N$-dimensional Einstein-Hilbert action
\be
S^{(N)}={1\over 16\pi l_{pl}^{N-2}}\int d^Ny\,\sqrt{-g}\,R^{(N)}(g)\,.
\label{l9a}
\ee
As we mentioned above for spherically symmetric configurations
\be
ds^2_N=\gamma_{\mu\nu}(x)\,dx^\mu dx^\nu +
G[\phi(x)]d\Omega^2_{N-2}\,,\label{l9b}
\ee
Eq.\ (\ref{l9a}) can be cast in the form (\ref{l1}) upon integration on the
angular coordinates. Let us now choose a different ansatz for the 
$N$-dimensional metric,
\be
d\tilde s^2_N={\tilde\gamma_{\mu\nu}\over\Omega(\phi)}\,dx^\mu dx^\nu +
G[\phi(x)]d\Omega^2_{N-2}\,.\label{l9c}
\ee
This ansatz is related to the first one by the ``conformal'' redefinition of
the two-dimensional metric field
$\gamma_{\mu\nu}=\tilde\gamma_{\mu\nu}/\Omega(\phi)$. The dimensionally reduced
action which is obtained by imposing the ansatz (\ref{l9c}) is still of the
form (\ref{l1}), where $\tilde{\cal V}(\phi)$ and $\tilde{\cal W}(\phi)$ are
related to ${\cal V}(\phi)$ and ${\cal W}(\phi)$ by Eq.\ (\ref{l6}). Note that
from the $N$-dimensional point of view the two-dimensional Weyl symmetry
(\ref{l6}) is just a field redefinition. Although the local properties of the
metrics $\gamma_{\mu\nu}$ and $\tilde\gamma_{\mu\nu}$ are different, the {\it
physical} properties of the $N$-dimensional system must be independent from the
ansatz that has been chosen, i.e.\ they do not depend on the two-dimensional
conformal frame which is being used. Therefore, only quantities which are
invariant under the symmetry (\ref{l6}) make physically
sense.\footnote{Actually, the ADM mass, the temperature, and the flux of
Hawking radiation turn out to be invariant under the Weyl transformation,
Eq.\ (\ref{l6}) \cite{Cadoni}.} 

Now, let us turn to the quantum theory. Two-dimensional pure dilaton gravity is
a general covariant, constrained, theory which is invariant under coordinate
reparametrization. The theory possesses two degrees of freedom (the dilaton and
a single gravitational degree of freedom that can be identified with the
conformal factor of the metric) and two constraints, so it is actually a
topological theory with no propagating degrees of freedom. Moreover, the
constraints can be solved and the central term can be made vanishing by a
suitable choice of the vacuum \cite{BJL}. The whole physical content of the
theory is given by the gauge invariant observables of the system. Because of
the topological nature of two-dimensional dilaton gravity, the observables
coincide with the conserved charges.

For theories described by the action (1) we have the single gauge invariant
quantity (see \cite{Cavaglia-PRD},\cite{LGK},\cite{LK}-\cite{CU} and 
references therein)
\be
M=N(\phi)-{\cal W}(\phi)(\nabla\phi)^2\,,~~~~N(\phi)=\int^\phi
d\phi'[{\cal W}(\phi'){\cal V}(\phi')]\,.
\label{l10}
\ee
The quantity $M$ is gauge invariant and locally conserved. Apart from a
constant normalization factor, for asymptotically flat geometries $M$ concides
on-shell with the ADM mass of the system. Moreover, $M$ is classically
invariant under the Weyl symmetry (\ref{l6}) \cite{Cadoni}. This can be proved
by direct checking or by noticing that the dilaton action (\ref{l1}) can be 
rewritten as a function of $M$ and $\phi$ as \cite{Cavaglia-PRD}
\be
S_{DG}=\int_\Sigma d^2x\,\sqrt{-g}\,{\nabla_\mu\phi\nabla^\mu M
\over N(\phi)-M}+\hbox{surface terms}\,.
\label{l11}
\ee
Since Eq.\ (\ref{l1}) is invariant under the transformation (\ref{l6}), and both
$\phi$ and the Weyl combination $\sqrt{-g}g^{\mu\nu}$ are Weyl invariant, $M$
must necessarily be invariant under Eq.\ (\ref{l6}).

The quantity $M$ is the only conserved charge of the theory and must determine
completely the latter. Indeed, solving the constraints the effective gauge
fixed action on the constraint shell ($\pi_\phi=0$, $M'=0$) is
\cite{Cavaglia-PRD,Kuchar}
\be
S_{eff}=\int d\tau\left[\frac{dm}{d\tau} 
p_m-m\right]\,,
\label{l12}
\ee
where $m=M|_{boundary}$, $p_m$ is the conjugate momentum of $m$, and $\tau$ is
the proper time on the boundary. In the classical regime the physical content
is completely determined by the value of $m$. In the quantum regime, the
Hilbert state of the theory is completely determined by the eigenstates of the
operator $\hat m$ \cite{Cavaglia-PRD,Kuchar}. This is the quantum
generalization of the no-hair theorem for a classical spherically symmetric
black hole in vacuo: A state is determined uniquely by the locally conserved
charge (the ADM mass) of the system. Since the quantity $M$ is invariant under
the transformation (\ref{l6}), quantum two-dimensional dilaton gravity do not
depend on the particular Weyl frame that has been chosen: Different frames lead
to the same $\hat m$. Let us stress that the action (\ref{l1}) can be cast in
the form (\ref{l12}) for any choice of $\Omega(\phi)$. Therefore, since $M$ is
invariant under the symmetry (\ref{l6}), and determines completely the Hilbert
space of the theory, the latter do not depend on the Weyl frame. A possible
objection to this statement could be that $\hat m$ is a rather special operator
and that, in general, operators corresponding to quantities which are not
classically invariant under the transformation (\ref{l6}) are affected by the
choice of the Weyl frame. Consequently, the quantum theory itself should depend
on the Weyl gauge fixing. While the first part of this objection is correct,
the conclusion is not. Indeed, since we are dealing with a constrained theory,
only gauge invariant operators make sense. Since $\hat m$ is the only gauge and
Weyl invariant operator of the theory, any operator which is not Weyl invariant
is necessarily not gauge invariant, so it does not have physical
interpretation. This conclusion is also obtained through a different approach
to dilaton gravity which has been worked in detail for the (matterless) CGHS
model \cite{CDF-PLB}. The essence of this approach is that the CGHS model
(\ref{l5}) can be rewritten in terms of a couple of free fields \cite{BJL}
which are pure gauge. Once more, the only physical quantity of the theory
coincides with the zero mode of the gauge and Weyl invariant mass operator.

In this letter we have seen that both classical and quantum pure
two-dimensional dilaton gravity are unaffected by Weyl transformations. For the
classical theory we have shown that Weyl transformations define a symmetry of
the system: The equations of motion are invariant under Weyl transformations.
The choice of the Weyl frame is analogous to a choice of gauge fixing. The
quantum theory of two-dimensional dilaton gravity is also {\it physically
invariant} under Weyl transformations, in the sense that the Hilbert space is
completely determined by the eigenstates of a (single) gauge and Weyl invariant
observable. Let us finally stress that these results do not hold for
matter-coupled two-dimensional dilaton gravity. For instance, if we couple the
Polyakov action to the model (\ref{l1}) the resulting theory is not
topological, the constraints cannot be solved, quantum anomalies appear and
Weyl invariance is generally lost\footnote{The quantization of the
matter-coupled model has been thoroughly investigated in the literature for the
CGHS case and several quantization approaches have been proposed to deal with
anomalies. (See \cite{BJL,ASJ} and references therein.)}. Therefore, Weyl
transformations are likely to play a very different role in matter-coupled
dilaton gravity.

\medskip\noindent
Acknowledgements. The author is very grateful to Stanley Deser and Roman Jackiw
for interesting discussions and useful comments about the content of this
paper. Some of this work was done while visiting Tufts University. The author
thanks Alex Vilenkin for his kind hospitality.

\end{document}